\input harvmac
\input epsf
\noblackbox


\overfullrule=0pt

\def\IP{{\bf P}}

\def\bfone{\relax{\rm 1\kern-.35em 1}}
\def\inbar{\vrule height1.5ex width.4pt depth0pt}
\def\IC{\relax\,\hbox{$\inbar\kern-.3em{\mss C}$}}
\def\ID{\relax{\rm I\kern-.18em D}}
\def\IF{\relax{\rm I\kern-.18em F}}
\def\IH{\relax{\rm I\kern-.18em H}}
\def\II{\relax{\rm I\kern-.17em I}}
\def\IN{\relax{\rm I\kern-.18em N}}
\def\IQ{\relax\,\hbox{$\inbar\kern-.3em{\rm Q}$}}
\def\us#1{\underline{#1}}
\def\IR{\relax{\rm I\kern-.18em R}}
\font\cmss=cmss10 \font\cmsss=cmss10 at 7pt
\def\ZZ{\relax\ifmmode\mathchoice
{\hbox{\cmss Z\kern-.4em Z}}{\hbox{\cmss Z\kern-.4em Z}}
{\lower.9pt\hbox{\cmsss Z\kern-.4em Z}}
{\lower1.2pt\hbox{\cmsss Z\kern-.4em Z}}\else{\cmss Z\kern-.4em
Z}\fi}

 \def\c{\gamma}

\def\cF{{\cal F}}

\def\cL{{\cal L}} 
 \def\cO{{\cal O}}

\def\nup#1({Nucl.\ Phys.\ $\us {B#1}$\ (}
\def\plt#1({Phys.\ Lett.\ $\us  {B#1}$\ (}
\def\cmp#1({Comm.\ Math.\ Phys.\ $\us  {#1}$\ (}
\def\prp#1({Phys.\ Rep.\ $\us  {#1}$\ (}
\def\prl#1({Phys.\ Rev.\ Lett.\ $\us  {#1}$\ (}
\def\prv#1({Phys.\ Rev.\ $\us  {#1}$\ (}
\def\mpl#1({Mod.\ Phys.\ Let.\ $\us  {A#1}$\ (}
\def\ijmp#1({Int.\ J.\ Mod.\ Phys.\ $\us{A#1}$\ (}
\def\tit#1|{{\it #1},\ }

\def\Coe#1.#2.{{#1\over #2}}

\def\coe#1.#2.{\relax{\textstyle {#1 \over #2}}\displaystyle}

\def\t#1{{\theta_#1}}
\def\tx{{\theta_x}}
\def\ty{{\theta_y}}
\def\tb{{\theta_b}}
\def\tf{{\theta_f}}
\def\ra{{\rightarrow}}
\Title{ \vbox{\baselineskip12pt\hbox{hep-th/9609239}
\hbox{EFI-96-37}
\hbox{HUTP-96/A046}
\hbox{OSU-M-96-24}}}
{\vbox{
\centerline{Geometric Engineering of Quantum Field Theories}}}
\centerline{Sheldon Katz$^1$,
Albrecht Klemm$^2$ and Cumrun Vafa$^3$}
 \medskip
\centerline{$^1$ Department of Mathematics}
\centerline{Oklahoma State University}
\centerline{Stillwater, OK 74078, USA}
\medskip
\centerline{$^2$ Enrico Fermi Institute}
\centerline{University of Chicago}
\centerline{Chicago, IL 60637, USA}
\medskip
\centerline{$^3$ Lyman Laboratory of Physics}
\centerline{Harvard University}
\centerline{Cambridge, MA 02138, USA}
\medskip
\centerline{\tt katz@math.okstate.edu, aklemm@maxwell.uchicago.edu,
 vafa@string.harvard.edu}
\vskip 1cm
\noindent
Using the recent advances in our understanding
of non-perturbative aspects of type II strings
we show how non-trivial exact results for $N=2$ quantum
field theories can be reduced to T-dualities of string theory.
This is done by constructing a local geometric realization
of quantum field theories together with a local application of mirror
symmetry.  This construction is not based on
any duality conjecture and thus reduces non-trivial
quantum field theory results to much better
understood T-dualities of type II strings.  Moreover
it can be used in principle to construct in a systematic
way the vacuum structure for
arbitrary $N=2$ gauge theories with matter representations.

\Date{September 1996}

\newsec{Introduction}
In the past few years string duality has taught us many
important lessons about non-perturbative aspects of string theory.
In each case where there is a reasonable duality conjecture, it
often requires certain non-perturbative corrections or non-trivial
solitonic spectrum for the duality to work.  As an example,
type IIA ($K3$) duality with heterotic ($T^4$) requires
that the $ADE$ singularities of $K3$ lead to enhanced gauge symmetry
of $ADE$ type \ref\witdiv{E. Witten,
{\sl String Theory Dynamics in Various Dimensions},
\nup443(1995) 85, hep-th/9503124,
{\sl Some Comments
On String Dynamics}, Proc. String's 95, hep-th/9507121}.
  This has been subsequently verified by realizing that $2$-branes
of type IIA wrapped around vanishing 2-cycles
lead to precisely the missing states expected for gauge symmetry enhancement
(generalizing the observation of \ref\strom{A. Strominger, \nup452(1995) 96,
hep-th/9505105}\ref\gms{B. Greene, D. Morrison, A. Strominger,
\nup451(1995) 109,
hep-th/9504145.}).
This has been analyzed in detail in
\ref\bsvi{M. Bershadsky, V. Sadov and C. Vafa,\nup463(1996) 398,
hep-th/9510225.}\ref\bsvii{M. Bershadsky, V. Sadov and C.
Vafa, \nup463(1996) 420,
hep-th/9511222.}.

In retrospect we do not need the postulate of duality to derive
the existence of enhanced gauge symmetry, even though that was helpful
in raising the question of physical interpretation of ADE singularity
of type IIA compactification on $K3$.  This is the case
with many similar examples and thus we can actually state that
independently of string duality conjectures, which
despite massive evidence to support them are
still conjectural, the existence
of solitons and their properties are on much firmer grounds and we can
claim to have {\it understood} how to deal with them in string theory.
This has become much firmer thanks to the very simple
description of D-branes as the relevant source for RR charge
\ref\pol{J. Polchinski,
{\sl  Dirichlet  Branes and Ramond-Ramond Charges},
\prl75(1995) 4724, hep-th/9510017}.
A manifestation of this understanding of solitonic objects in string theory
 is the ability to count the appropriate
D-brane states and account for extremal black hole entropy
which is not
predicted by any known string duality \ref\sv{A. Strominger and C. Vafa,
{\sl Phys. Rev. Lett.} 77 (1996) 2368}.

Certain non-perturbative results for quantum field theories
has been derived using string dualities.  A particular
class of such theories involve $N=2$ supersymmetric
theories in $d=4$ where type II on CY 3-fold is dual
to heterotic on $K3\times T^2$
\ref\kv{S. Kachru and C. Vafa, \nup450(1995) 69,
hep-th/9505105.}\ref\FHSV{S. Ferrara, J. Harvey, A. Strominger and C. Vafa,
\plt361(1995) 59 , hep-th/9505162.}\ref\ogen{
V.\ Kaplunovsky, J.\ Louis, and S.\ Theisen, \plt357 (1995) 71;
A.\ Klemm, W.\ Lerche and P.\ Mayr, \plt357 (1995) 313;
C.\ Vafa and E.\ Witten, preprint HUTP-95-A023;
hep-th/9507050; I.\ Antoniadis, E.\ Gava, K.\ Narain and
T.\ Taylor, \nup455 (1995) 109,
B. Lian and S.T. Yau,
{\sl Mirror Maps, Modular Relations and Hypergeometric Series I,II},
hep-th/9507151, hep-th 9507153;P.\ Aspinwall and J.\ Louis, \plt369 (1996) 233;
I.\ Antoniadis, S.\ Ferrara and T.\ Taylor,\nup 460(1996) 489;
G. Curio, \plt366 (1996)  131,\plt366 (1996)  78;
G.\ Lopes Cardoso, G. Curio, D. L\"ust and T. Mohaupt,
{\sl Instanton Numbers and Exchange Symmetries in $N=2$ Dual String
Pairs}, hep-th/9603108}.
These dualities fit well with our understanding
of how gauge symmetry arises in type II theories
due to the observation \ref\klm{A. Klemm, W. Lerche and
P. Mayr, as cited in \ogen}\
that the main examples in \kv\ involve $K3$ fibered
CY and the adiabatic argument \ref\vw{C. Vafa and E. Witten, as cited
in \ogen}\
reduces the mechanism of the gauge symmetry enhancement
to the case of $K3$.  This aspect of it is
in fact emphasized in deriving the field theory
consequences of these dualities \ref\kklmv{S. Kachru,
A. Klemm, W. Lerche, P. Mayr and C. Vafa,
hep-th/9508155.}\ref\klmvw{A. Klemm, W. Lerche, P. Mayr, C. Vafa, N. Warner,
hep-th/9504034.}.

The question we wish to address in this paper is whether
we can {\it derive} non-trivial field theory
results directly as a consequence of the recently acquired
deeper understanding of string theory dynamics, rather than
as a result of a consequence of a duality conjecture.
If so we can claim to {\it understand} non-trivial
results in field theory simply based on the existence
of string theory and its established
properties! As we shall see this will lead directly
not only to
a stringy confirmation of
 conjectured results in field theory but also to potentially
new results in field theory.
An interesting arena to test these ideas is in the case
of $N=2$ supersymmetric theories in $d=4$, which
is what we will focus on in this paper, even though
we believe our methods should be generalizable to other
cases and in particular to $N=1$ theories in $d=4$
(for some results in this
directions see in particular
\ref\ksi{S. Kachru and E. Silverstein, hep-th/9608194}).

In an $N=2$
field theory one starts with a gauge group $G$, with matter
in some representation $\oplus R_i$ of $G$. The mass for matter
can be incorporated by hand, or more naturally by considering
$G$ to have some $U(1)$ factors under which the $R_i$ are charged,
and by going to the Coulomb phases of the $U(1)$ factors.

Our basic strategy is to find a situation in type IIA theory
where exactly the same gauge theory and matter content arise.
More precisely, since the gauge symmetries and matter arise
near a singular limit of type IIA compactification, and sending
$M_{planck}\rightarrow \infty$ corresponds to studying the local
part of the singularity, we do not have to find a {\it global}
description of type IIA compactification; a {\it local} description
of the singularity suffices.
$N=2$ supersymmetry in $d=4$ arises for type IIA compactification on
Calabi-Yau threefold.  So we are looking for a local model
where a Calabi-Yau singularity gives rise to the requisite
gauge group $G$ with the matter content $R_i$.  As for the
gauge group, given that it arises in 6 dimensions once we have
$ADE$ singularities, if we fiber that over some surface
we could easily obtain $ADE$ gauge symmetry in 4 dimensions
\ref\aspi{P. Aspinwall, \plt357(1995) 329,\plt371(1996) 231,}\bsvi \ref\kmp{S.
Katz, D. Morrison and R. Plesser,hep-th/9601108}\ref\km{A. Klemm and P. Mayr,
hep-th/9601014},
or by using outer automorphisms \ref\asg{P. Aspinwall and
M. Gross, hep-th/9602118}\
we can even realize non-simply laced gauge groups in
4 dimensions \asg\ref\sixau{
M. Bershadsky, K. Intriligator,S. Kachru,
 D. Morrison, V. Sadov and C. Vafa, hep-th/9605200.}.
In particular if we have a genus $g$ curve with $ADE$ singularity
one expects to have $ADE$ gauge symmetry in 4 dimensions with $g$
adjoint hypermultiplets \kmp \km.
If we are interested in asymptotically free theories we would want
to have the genus of the curve be $g\leq 1$.  The case $g=1$ will
lead to $N=4$ spectrum.  So we will consider the $g=0$ case.

Other matter representations can also be obtained
from more intricate singularities
\bsvi\ref\bkkm{P. Berglund, S. Katz, A. Klemm and P. Mayr,
hep-th/9601108}\sixau
\ref\kv{S. Katz and C. Vafa, hep-th/9606086}.
In particular
if we have an $ADE$ singularity over a surface and at some points along
the surface the singularity is enhanced to a higher one, then this
is equivalent to some matter localized at those points \kv .  For example
if we have an $A_{n-1}$ singularity in the fiber which is enhanced
to $A_{n}$ at $k$ points on the surface we end up getting
$SU(n)\times U(1)$ gauge symmetry with $k$ hypermultiplets in the
fundamental of $SU(n)$ charged under $U(1)$ \kv. This spectrum
can be understood locally as breaking of the adjoint of $SU(n+1)$
to $SU(n)\times U(1)$ at $k$ points on the surface, each leading
to a hypermultiplet.  Other examples of gauge groups with
various matter representations have been discussed in \kv .

The basic organization of this paper is as follows.
In section 2 we start our discussion of the setup for the
simple example of $SU(2)$ without matter and the limit
of turning off gravity.   In section 3
we discuss quantum correction (or lack thereof) in this context
and the relation of worldsheet instantons to pointlike
spacetime instantons.
In section 4 we review some facts about mirror symmetry applied
in a local sense.  Even though mirror symmetry is well
known, the local application we have in mind has not been fully
developed
(except in some special cases, e.g. some universal
properties of CY-threefolds admitting flops  were discussed in
\ref\wittenlg{E. Witten, \nup403(1993) 159}\ref\agm{P. Aspinwall,
B. Greene, D. Morrison, \nup420(1994) 184}, other examples appeared
in \ref\kmv{A. Klemm, P. Mayr and C. Vafa, hep-th/9607139}\agm
); we remedy this in
section 4.  In section 5 we return to the case of $SU(2)$ without
matter and show how we can use mirror symmetry results
and arrive at the field theory results of Seiberg and Witten
\ref\swi{N. Seiberg and E. Witten,
\nup426(1994) 19, hep-th/9407087} \ref\swii{N. Seiberg and E. Witten,
\nup431(1994) 484, hep-th/9408099}.
In section 6 we discuss some generalizations
including $SU(n)$ without matter and some examples with matter.
We hope to return to a more complete list of examples in a forthcoming
work \ref\kkv{A. Klemm, S. Katz and C. Vafa, work in progress.}.

\newsec{$SU(2)$ without matter: an illustrative example}

To obtain an $SU(2)$ gauge symmetry we need
an ${\bf A}_1$ singularity in $K3$.  This means that we need
a vanishing 2-sphere
${\bf P}^1$.  The $W^{\pm}$ will correspond to 2-branes wrapped
around ${\bf P}^1$ with two opposite orientations.  The mass
of $W^{\pm}$ is proportional to the area of ${\bf P}^1$.
To obtain an $N=2$ theory in four dimension we need to fiber
this over another 2-sphere ${\bf P}^1$.  The number of
ways this can be done is parameterized by an integer $n$ corresponding
to the Hirzebruch surfaces ${\bf F}_n$. The simplest case is
${\bf F}_0={\bf P}^1\times {\bf P}^1$.  All the ${\bf F}_n$ in the limit
of vanishing K\"ahler class of the fiber $P^1$ are expected to give
rise to $SU(2)$ gauge symmetry in $d=4$ with $N=2$ supersymmetry
without any matter.  Note that $1/g^2$
is proportional to the area of the base ${\bf P}^1$ because
of the compactification from $6$ to $4$, where $g$ is the bare gauge coupling
constant defined at the string scale.

\subsec{The $M_{planck}\rightarrow \infty$ limit}

To decouple gravitational and stringy effects and recover
pure field theory results all we have to do
is to consider a limit discussed in \kklmv .
In particular the bare coupling constant $g$ at the string scale
should go to zero if the string scale is being pushed
to infinity, simply by running of the gauge coupling
constant and asymptotic freedom.  Thus
we take a limit $t_b\rightarrow \infty$ where $t_b$ denotes
the K\"ahler class of the base.
  On the other hand
in order for us to have a finite $W^{\pm}$ mass in this limit
we have to take the area of the fiber ${\bf P}^1$ which in string units
corresponds to $mass\, (W^{\pm}/M_{string})\rightarrow 0$,
we thus have $t_f\rightarrow 0$, where $t_f$
denotes the K\"ahler class of the fiber.  Clearly the two limits
are not unrelated.  In particular in the weak coupling regime
the running of the coupling constant is dominated by
$${1\over g^2}\sim {\rm log}\, {M_W\over \Lambda}$$
which implies that we should consider the limit where
$$t_{b}\sim -{\rm const.}\ {\rm log}\, {t_f}$$
We can be more precise.  The gauge theory instanton number $n$ will
be weighted with ${\rm exp}(-n/g^2)$.  On the other hand from field theory
we know that each power of instanton is accompanied by $1/a^4$
where $a$ is the Cartan expectation value of the $SU(2)$ adjoint,
which is proportional to $t_f$.  Putting these together we
thus look for the limit
$${\rm exp}(-1/g^2)={\rm exp}(-t_b)\sim \epsilon^4 \Lambda^4$$
\eqn\limi{t_f\sim  \epsilon a}
as $\epsilon \rightarrow 0$.
In the following we often set the scale $\Lambda =1$.

\newsec{Quantum corrections}
So far our description of the Coulomb branch of $N=2$ theories
in terms of K\"ahler classes of Calabi-Yau
manifolds in type IIA strings has been purely classical.  In the
context of strings we can have two types of corrections to the
classical result:  worldsheet corrections and quantum string corrections.
As is well known the K\"ahler moduli of Calabi-Yau compactifications
{\it do} receive corrections due to worldsheet instantons.
However thanks to mirror symmetry this is well under control
and can be computed exactly, at least for a wide range of cases.
As far as stringy quantum corrections to K\"ahler moduli are
concerned they are absent, since the coupling constant
field of type IIA is in a hypermultiplet and the K\"ahler moduli belong
to vector multiplets and thus do not talk with each other
\ref\nonre{B. de Wit, P.G. Lauwers,
R. Philippe, S.Q. Su, A. van Proeyen, \plt134(1984) 37;
B. de Wit, A. van Proeyen, \nup245(1984) 89; J.P. Derendinger,
S. Ferrara, A. Masiero,
A. van Proeyen, \plt140(1984) 307; B. de Wit, P.G. Lauwers,
A. van Proeyen, \nup255(1985) 569; E. Cremmer, C. Kounnas, A. van Proeyen,
J.P. Derendinger, S. Ferrara, B. de Wit, L. Girardello,
\nup250(1985) 385}\ref\bsu{N. Berkovits and W. Siegel, \nup462(1996) 213},
and so we can
take the limit of weak string coupling to argue that string tree
level should be exact.  Thus the only corrections to the
classical description we need to worry about are the worldsheet
instantons.

In order to interpret the worldsheet corrections we need
to recall some basic facts about compactifications of type IIA
on $K3$.  In type IIA
compactification on $K3$ there is an important field equation induced in
6 dimensions
\ref\vwt{C. Vafa and E. Witten, \nup447(1995) 261}
\eqn\ins{d*({\rm exp }(-2\phi) )H=tr R\wedge R-tr F\wedge F}
The case we are considering in 4 dimensions is obtained by
considering a fibration of this case over another ${\bf P}^1$.
Let us consider a worldsheet instanton wrapped $n$ times around the
base ${\bf P}^1$ at a given point $x$ in the uncompactified
spacetime.  If we surround the point $x$ by a 3-sphere in spacetime
and integrate, by definition of the wrapping of the sphere we learn that
$$\int_{S^3}*({\rm exp }(-2\phi)) H =n$$
Because of \ins\ this means that we should identify worldsheet
instantons which wrap $n$ times around the base as
corresponding to point-like gauge theory instantons, with instanton number $n$.
 Note that we are
ignoring the $\int R\wedge R $ term above, because
in the limit of $M_{string}\rightarrow \infty$ they are
not relevant.  That sometimes the worldsheet instantons play the role of gauge
theory instantons was anticipated based on string dualities
\ref\duff{M. Duff and R. Khuri, \nup411(1994) 473,
C. M. Hull and P.K Townsend, \nup438(1995) 109}\witdiv .
Our main point in repeating this
here is to emphasize that we do not need any input from
string duality to make this link. It can be understood purely in
the framework of perturbative type IIA string theory.

\subsec{Prepotential of $N=2$ and the worldsheet instantons}
As is well known the contribution to the prepotential
of $N=2$ theories is related to the number of
worldsheet instantons which itself can be computed
using mirror symmetry.  Let $C_i$ denote a basis for
$H_2$ and let $d_{n_i}$ denote
the number of primitive instantons in the K\"ahler class
denoted by $\sum n_iC_i$, then the prepotential
${\cal F}$, which is a
function of the K\"ahler parameters $t_i$, satisfies
\ref\cand{P. Candelas, X. de la Ossa, P. Green and
L. Parkes, \nup359(1991) 21}
$$\partial^3_{uvw}{\cal F}
=\sum_{n_i}n_un_vn_w d_{n_i}{\prod_i q_i^{n_i}
 \over 1-\prod_i q_i^{n_i}}$$
where $q_i={\rm exp}(-t_i)$.
Once we know what $d_{n_i}$ are and what the limit
of turning off gravity effects are, as discussed above,
we can thus reproduce the field theory results directly. This
has the advantage that will allow us not only to identify
the global aspects of the Coulomb branch, but also
allows us to isolate individual contributions of worldsheet
instantons to the gauge field instantons.
In particular if we consider the contribution to
${\cal F}$ coming from worldsheet instantons wrapping
$n$ times around the base we can see how they
would lead to instanton number $n$ corrections to the prepotential
in the gauge theory system.

\newsec{Local mirror symmetry}
In this section we explain our construction of local mirror
symmetry. We consider a local neighborhood $X$ of a surface
$S$ in a Calabi-Yau threefold. The local geometry will be enough
to describe the type IIA string theory compactified on the
Calabi-Yau threefold $M$ in certain physically interesting limits,
and will be independent of the choice of Calabi-Yau containing
$X$ in this same limit.
The type IIA theory will depend on the K\"ahler parameter
of the local geometry, which we call local $A$ model,
in a very complicated way due to worldsheet instantons corrections.
We will therefore introduce a mirror description of the local geometry,
which we refer to as local $B$ model. Physical quantities
of interest in this paper are
exactly given in terms of integrals over a meromorphic form
in the mirror geometry. In particular we are interested in the
cases when the local model for $S$ contains vanishing cycles
so that in the limit particle spectra with interesting gauge symmetries
and matter content become massless.

Our construction can be given equivalently in the context of the
linear sigma model \wittenlg
\ref\mp{D. Morrison and
R. Plesser, \nup440(1995) 279, hep-th/9412236,
{\sl Towards Mirror Symmetry as
Duality For Two-Dimensional Abelian Gauge Theories},
hep-th/9508107 }\ or in the context of toric geometry
\ref\bbb{V. Batyrev, Duke Math. Journal 69
(1993) 349, V. Batyrev, Journal Alg. Geom. 3 (1994) 493,
V. Batyrev, {\sl Quantum Cohomology Rings of Toric Manifolds},
preprint 1992, V. Batyrev and A. Borisov, alg-geom/9412017,
alg-geom/9509009}.

\subsec{The local $A$ model}
We will construct the local model for the K\"ahler geometry
as gauged $N=2$  two dimensional linear sigma model
\wittenlg \ with
$n+3$ chiral multiplets $X_i$ which are charged under a $G=U(1)^n$
gauge group. The charges of the fields $X_i$ are grouped into charge
vectors $v^{(k)}=(q_1^{(k)},\ldots, q_{n+3}^{(k)})$. Non anomalous
$R$-symmetry of the field theory is equivalent to triviality of the canonical
bundle
in the neighborhood $X$ and implies $\sum_{i=1}^{n+3} q^{(k)}_i=0$.
To get the space of classical vacua we analyze the zero locus
of the scalar potential $U$. The latter is in absence of a
superpotential just given by the $D$ terms
\eqn\scalarpotetial{U= \sum_{k=1}^n {1\over 2 e_k^2} D_k^2, \ \
{\rm with} \ \ D_k=-e_k^2(\sum q_{i}^{(k)} |x_i|-r_k),}
where the $r_k$ are to be identified with the K\"ahler parameters, and
describes in our case a three dimensional variety.

To illustrate this we consider a theory with $U(1)\times U(1)$ gauge group
and charges $v^{(1)}=(1,1,-2,0,0)$, $v^{(2)}=(0,0,1,-2,1)$.
The scalar potential reads
$$U={e_1^2\over2}(|x_1|^2+|x_2|^2-2|x_3|^2-r_1)^2+
{e_2^2\over2}(|x_3|^2-2|x_4|^2+|x_5|^2-r_2)^2$$
In the phase where $r_1$ and
$r_2$ are both positive,\foot{The $U(1)$ generators were chosen so that the
phase we are interested in corresponds to requiring $r_1$ and $r_2$ to be
positive---this is tantamount to choosing a basis for the Mori cone, as we will
see.} we see that for our vacua we cannot have
$x_1=x_2=0$ or $x_3=x_5=0$.  The set of these fields modulo gauge equivalence
is parameterized by a local threefold $X$ with trivial canonical bundle.
The space $X$ contains a surface $S$ defined by $x_4=0$, which is
in fact the Hirzebruch surface $F_2$ that is a ruled surface
over ${\bf P}^1$ with fiber ${\bf P}^1$.  Thus the non-compact
Calabi-Yau manifold $X$ has a compact part which we identify
with $F_2$ and the total space $X$ can be identified with
$F_2$ together with the normal bundle on it (which is identified with
the canonical line bundle for $F_2$ given here by the $x_4$ direction).
Our model resolves a
curve of $A_1$ singularity parameterized by ${\bf P}^1$; the
fibers of $F_2$ are the vanishing cycles. The well-known cohomology of
$F_2$ is generated by the class $s$ of a section with $s^2=-2$ and
the class $f$ of a fiber.  The other intersection numbers are
$s\cdot f=1$ and $f^2=0$.  There is another section $H=s+2f$ which
is disjoint from $s$. The section $s$
itself is  defined by $x_3=x_4=0$.  The curves $x_1=x_4=0$ and $x_2=x_4=0$
are fibers $f$ of the Hirzebruch surface.  The section $H$ is identified with
$x_4=x_5=0$.  The divisors $x_i=0$ for $i=1,2,3,5$ are noncompact divisors
in $X$ which intersect $S$ in the respective curves $f,f,s,H$.  For this
reason, we will sometimes refer to these divisors as $f,f,s,H$.  The divisor
$S$ restricts to $S$ itself as the canonical class $K=K_S$ by the
adjunction formula.

Starting from the geometry we can find the correct charge identification
as follows. We identify K\"ahler classes on $F_2$ with curves, and choose
as a basis the classes of $s$ and fiber $f$ (which generate
the Mori cone of $F_2$).  Identifying each of these with a $U(1)$, we have
to find the charges of the fields under $G$.  The fields $x_i$ are identified
with divisors $x_i=0$ on our local model $X$.  The charge of a field
under a $U(1)$
is identified with the intersection number of the divisor associated to the
field with the curve representing the $U(1)$.
We choose divisors in the order $f,f,s,K,H$.  Using the formula
$K=-2s-4f$, the charges under the first $U(1)$ are identified with the
intersection numbers of the 5~divisors with $s$, while the charges
under the second $U(1)$ are identified with the
intersection numbers of the 5~divisors with $f$.

The identical situation can be described and clarified somewhat
by toric geometry.  We seek a
3~dimensional toric variety $X$ containing $F_2$ as a divisor.  As a toric
variety, $F_2$ has 4~edges; there must also
be a fifth edge associated to the divisor $F_2$ in $X$.  This is why
5~fields were needed in our model.  The calculation concludes as before.
The charge vectors $(1,1,2,0,1)$ and $(0,0,1,-2,1)$ are identified with
the Mori vectors of $F_2$ in this context.

Now consider a local $A$ model arising from compactifying an appropriate
string theory on a Calabi-Yau hypersurface in a toric variety $V$ associated
to a reflexive polyhedron $\Delta$. Such a theory has a linear sigma
model analogue which has $k$ additional fields\foot{Note that it also
contains an additional auxiliary field called $P$ in \wittenlg ,
which is necessary to implement the constraint.}  $X_i$ and $k-1$ additional
$U(1)$ gauge factors but since it includes a superpotential, which leads to
a constraint, its vacuum configuration is of the same dimension.
Suppose that our local model $X$
considered above is contained in $V$
which in particular means that the
fields $X_i$ are part of the fields for the compact linear sigma model
as well.  Put in the toric language \bbb,  we suppose that
the fan of $X$ is generated by some of the integral points of $\Delta$.
In terms of the linear sigma model, this means that after restricting
to a subset of 5 of the fields of the full theory, two charge vectors
can be found which coincide with the vectors $(1,1,-2,0,0),(0,0,1,-2,1)$.

\subsec{The local $B$ model}

The data of the local $B$ model as we will see
gives rise to a complex variety of dimension 2 less than $X$
(Riemann surface for the case of CY 3-folds) and
a meromorphic form $\Omega$ on it.
Physical quantities in the appropriate
limit of the Calabi-Yau threefold are exactly given by periods
of $\Omega$ on this Riemann surface
which we denote by $\hat S$.  It turns out that $\hat S$
is obtained as a hypersurface in the projectivization of
 a more natural 3-fold
mirror which we denote by $\hat X$.

We start by constructing $\hat X$ following a method of
Batyrev-Borisov \bbb, which has an equivalent formulation in terms
of the linear $\sigma$-model \wittenlg \mp .

We introduce variables $y_i$ to each field in the linear sigma model.
(or equivalently for each edge of the toric fan).
Then the $y_i$ may be interpreted as
coordinates on $\hat{X}$, and for each charge vector
$v^{(i)}=(q^{(i)}_1,\ldots,q^{(i)}_r)$, the relation
\eqn\localmirror{\prod_{q^{(i)}_j>0}y_j^{q^{(i)}_j}=
\prod_{q^{(i)}_j<0}y_j^{-q^{(i)}_j}}
is satisfied. For example, the quintic has a single charge vector
$(-5,1,1,1,1,1)$, and the quintic mirror can be coordinatized by the
forms
$$(y_0,\ldots,y_5)=(x_1\cdots x_5,x_1^5,\ldots x_5^5).$$
  These
satisfy the relation $y_0^5=y_1\cdots y_5$ as claimed. In our case
we get $n$ equations on the $n+3$ fields $y_i$ leaving us with
local coordinates $(t_1,t_2,t_3)$. As we will see below the relevant
aspect of
the local mirror geometry is given by a projectivisation
of the coordinates $(t_1,t_2,t_3)$.  We define $\hat S$ therefore as the
complex one dimensional hypersurface
obtained in projectivizing all the $y_i$'s  by
\eqn\constraint{P=\sum_{i=1}^{n+3} a_i {y_i(t)}=0}

E.g. in the case of our $F_2$ model, we get equations
$$y_1y_2=y_3^2,\ y_3y_5=y_4^2.$$

The equations can be solved by putting
\eqn\locmirparam{(y_1,\ldots,y_5)=(z,s^4/z,s^2,st,t^2).}
Thus $(z,t)$ (defined by projectivizing and setting $s=1$)
can be taken as local coordinates on $\hat S$ subject
to the condition
\eqn\swp{P=a_1 z+a_2 {1\over z} + a_3 + a_4 t+a_5 t^2=0.}

In a compact situation the quantum corrected K\"ahler
moduli of a compact Calabi-Yau $M$
are given by the periods of the holomorphic 3-form on the
mirror $\hat M$.  In our case we have a local
part of $M$ denoted by $X$ and we are looking
for the worldsheet quantum corrections to it.
We have already argued that the local model
for the mirror naturally contains a Riemann
surface.  It is thus natural to expect
that the relevant period integrals of $\hat M$
are reduced in this limit to period integrals
of an appropriate 1-form on the Riemann surface.  As we
will now show this is indeed the case.

The relevant periods of $\hat M$ are governed
by the Picard-Fuchs equations which can be derived directly
from the charge vectors of the local $A$-model. In the simplest
cases they are given by
\eqn\pf{
\prod_{q_i^{(k)}>0}\left({\partial\over \partial a_i}\right)^{q_i^{(k)}}=
\ \ \ \ \prod_{q_i^{(k)}<0}
\left({\partial\over \partial a_i}\right)^{-q_i^{(k)}}}
in terms of the charge vectors, where the $a_i$ are the coefficients
of the fields in the superpotential of the gauged linear
$\sigma$ model of the global $B$-model. As $P$ is actually a
restriction of this superpotential they are the same coefficients
as the ones appearing in \constraint\ .  In the present context
we are giving only a subset of differential equations which
are governed by the compact mirror, simply because
we only are interested in the K\"ahler classes which control
the size of $X\subset M$.  This however implies that
no matter what the rest of the data which go into defining
the compact version $M$, its periods still satisfy the
PF equations \pf\ where the parameters control the K\"ahler
classes of $X$ in $M$.

As an example consider the $F_2$ case discussed above.
In this case the Picard-Fuchs equations
simplify\foot{We will use here the same conventions as in \ref\HKTY{
S. Hosono. A. Klemm, S. Theisen and S.-T. Yau, \cmp 167(1995) 301}} to
\eqn\ftwo{\eqalign{
\cL_1&=\tb^2-z_b(\tf-2 \tb)(\tf-2\tb-1)\cr
\cL_2&=\tf (\tf-2\tb)-z_f(2\tf)(2 \tf +1)}}
where $\theta_i=z_i\partial /\partial z_i$ and
 we have introduced the invariant coordinates
$z_b=a_1a_2/a_3^2,\ z_f=a_3a_5/a_4^2$ on the moduli
space of $\hat M$.   More generally we define one $z$
for each charge vector by the relation $z_k=\prod a_i^{q^{(k)}_i}$.
The crucial point is that
{\it these operators are independent of the choice of
$(M,\hat M)$ which contains our local $A$- and $B$- models\/}.

We are thus left with the task of finding a way of solving
these PF equations.  It is in this way that the Riemann
surface we constructed above becomes useful, as we will now see.
We will define a 1-form $\Omega$ on the Riemann surface whose
periods solve the relevant PF equations for the mirror.

   Note that after
solving the equation for constraints of $y_i$ we are left
with three independent generators $y_i$.  Moreover
since we are interested in the projectivization
we can consider only a pair of them by going to the
inhomogeneous coordinates.
Let $y_1$ and $y_2$ be
these two variables.   We define $\Omega$  as follows:
\eqn\oneform{\Omega=\log P  {d y_1\over y_1}\wedge {d y_{2}\over y_{2}}.}
Note that this form is well defined up to an addition independent
of the moduli.  This ambiguity in the shift is reflected by the
fact that in the systems we consider 1 is always a solution to the
PF equations.
This obviously fulfills the Picard-Fuchs equation \pf\ and is a function
of $z_i$\foot{The periods $F$ of interest are scale invariant, in
the sense that
$F(ta)=F(a)$ where $a=(a_1,\ldots, a_5)$.  This is already clear from
our discussion of the differential operators, since $F$ is a function of
$z_1,z_2$.  This freedom allows us to set $s=1$ in \locmirparam\ to get
$(z,1/z,1,t,t^2)$ which are already visible as the monomials which
describe the Seiberg-Witten $SU(2)$ curve.  We will make this more precise
in the next section.}.
 The derivatives of this form
\eqn\bunchofforms{a_i {\partial\over \partial a_i}\Omega=
{a_i y_i\over P } {d y_1\over y_1}\wedge {d y_{2}\over
y_{2}}}
give by performing the residue integral around $P=0$ forms
on the Riemnann surface which
are unambiguous and among which are the holomorphic ones.
This can be viewed, alternatively, as rigorizing the definition
of the form $\Omega $ given above.

Next we like to see that \oneform\ reduces to the meromorphic
one form on the general Riemann surfaces for $N=2$ gauge theories
we construct from toric data.

To see this we partially integrate one coordinate and then perform the
integration over ${d P \over P}$

\eqn\ttr{\eqalign{\int \log P {d y_1\over y_1}\wedge {d y_{2}\over
y_{2}} &= \int \log P d \log y_1\wedge
{d y_2\over y_2}\cr & =
-\int \log y_1 {d P\over P}\wedge{d y_2\over y_2}\cr & =
-\int \log y_1 {d y_2\over y_2}}}

\subsec{Other $F_n$ examples}
As discussed previously if we have any $F_n$ surface
sitting in the CY 3-fold, in a suitable limit where the fiber
shrinks and the base grows, we expect an $SU(2)$ gauge symmetry.
Above we showed how we can geometrically `engineer' the $F_2$
inside a Calabi-Yau.  Let us now illustrate this engineering
for two more cases $F_0={\bf P}^1\times{\bf P}^1$ and $F_1$.
In the next section we will check that all these cases
will lead to the same results in the field theory limit, as expected.

We first consider $S={\bf P}^1\times{\bf P}^1$.  There are two different
fibrations
in this case; we denote the respective fibers by $f_1$ and $f_2$.
The intersection numbers are $f_1^2=f_2^2=0,\ f_1\cdot f_2=1$. Realizing
$S$ as a toric variety, we have 4~edges, corresponding to divisors with
cohomology classes $f_1,f_1,f_2,f_2$ (the fibers over $0$ and $\infty$
for each fibration).  As in the $F_1$ case, there is another toric divisor
in the local model $X$, namely the divisor $S$, which restricts to
$K=K_S=-2f_1-2f_2$.  The generators of the
Mori cone are $f_1$ and $f_2$.  Ordering the divisors as
$K,f_2,f_2,f_1,f_1$, we get the charge or Mori vectors
$(-2,1,1,0,0),(-2,0,0,1,1)$ and the equations
$$\theta_{z_1}^2-z_1(2\theta_{z_1}+2\theta_{z_2})
  (2\theta_{z_1}+2\theta_{z_2}+1)=0,$$
\eqn\fzero{\theta_{z_2}^2-z_2(2\theta_{z_1}+2\theta_{z_2})
  (2\theta_{z_1}+2\theta_{z_2}+1)=0,}
with $z_1=a_2a_3/a_1^2$ and $z_2=a_4a_5/a_1^2$.

Our third example is $S=F_1$.  Its cohomology is generated by a section
$s$ with $s^2=-1$ and a fiber $f$.  The other intersection numbers
are $s\cdot f=1$ and $f^2=0$.  As a toric variety, $S$ has another
toric divisor which is another section $H$ in the class $s+f$.  In this
case, we have $K=-2s-3f$.  Choosing the divisors in the order
$K,H,f,s,f$ and identifying the $U(1)$ factors with the Mori generators
$s,f$, we compute the charge vectors to be
$(-1,0,1,-1,1),(-2,1,0,1,0)$.
We put $z_1=a_2a_4/(a_0a_3)$ and $z_2=a_1a_3/a_0^2$.
We derive the equations
$$\theta_{z_b}^2-z_b(-\theta_{z_b}-2\theta_{z_f})
         (-\theta_{z_b}+\theta_{z_f})=0,$$
\eqn\fone{(\theta_{z_f})(-\theta_{z_b}+\theta_{z_f})-
    z_f(-\theta_{z_b}-2\theta_{z_f})(-\theta_{z_1}-2\theta_{z_f}-1)=0.}

In the limit as the fibers of $F_0$, $F_1$ or $F_2$ shrink,
we get an $SU(2)$ enhancement of the gauge symmetry.
In the next section, we show from the local Picard-Fuchs
equations that they lead to the same physics in the limit, as expected.

\newsec{$SU(2)$ with no matter revisited}
We now return to the example we started in the paper
namely the case where we expect $SU(2)$ without any matter,
e.g. $F_0$ ($=\IP^1\times \IP^1$), $F_1$ and $F_2$.
To begin with, the prepotential will be a function
of two K\"ahler classes $t_b,t_f$ corresponding
to the base and fiber respectively.  Let
$d_{n,m}$ denote the number of primitive worldsheet instantons
wrapping $n$ times around the base and $m$ times around
the fiber.  Let us first
consider the prepotential in the absence
of gauge theory instantons.
 As discussed before
this should correspond to contributions from worldsheet
instantons with zero winding around the base.  The relevant
instanton numbers to compute are
$d_{0,m}$.   We have
\eqn\lincon{d_{0,1}=-2, \quad d_{0,i}=0,\ \  \forall i>1.}
This is easy to see.  We choose a point on the base
and wrap a curve around the fiber $\IP^1$.   There
is only one such $\IP^1$, so apart from the multicover
we have only the $d_{0,1}\not=0$.  Since we can choose
any point on the base to wrap around the $\IP^1$ we have
the base $\IP^1$ family of degree 1 maps.  The corresponding
`number of instantons' in this case is simply the
characteristic class on this moduli space  which is $c_1(\IP^1)=-2$.

We now simply consider the instanton expansion of the three point
coupling in which the three points of the sphere are mapped to
the divisor dual to the fiber, i.e. $\partial^3_{t_f} \cF$.
After integrating this w.r.t $t_f$ it should reproduce in the
 double scaling limit discussed in section (2.1)
exactly the running coupling of the field theory $\partial^2_a \cF =
-i\tau={4 \pi \over g^2}+2 \pi i\theta$.

Let us consider the contribution to the gauge coupling
constant which is encoded in the second derivative
of the prepotential, which is
$$\partial_{t_f}^2 \cF =
\sum_{n=0}^\infty \sum_{m=1 \atop k=1}^\infty d_{n,m} m^2
{q_b^{nk} q_f^{mk}\over k}.$$
If we now consider the perturbative limit, i.e. concentrate
on the terms in the above expansion with $n=0$, and use the
value of $d_{0,m}$ noted above we find
$$\partial_{t_f}^2 \cF_{pert}=-i\tau_{pert}=
-2\sum_{k=1}^{\infty}{q_f^k\over k}=2{\rm log}
(1-q_f).$$
As discussed in section 2, we need to consider the limit
$$q_f={\rm exp}(-t_f)\rightarrow {\rm exp}(-\epsilon a)\rightarrow 1-\epsilon
a$$
Thus we obtain in this limit we obtain the
running coupling constant $-i\tau_{pert}= 2 \log(a) +{\rm const.}+ O(\epsilon
)$.
That is, the logarithm of the field theory 1-loop diagram comes
from the single type holomorphic curve wrapping once the fiber plus
its multicover contributions.  Aspects of this correspondence
was already noted in \kmp\ in the context of gauge theories
with asymptotically non-free matter.
Much more generally for curves wrapping around
the components of the fibers which are sphere trees of $ADE$-type the
contributions from the single types of curves will likewise
resum to the perturbative part of the $N=2$ coupling constants
with $ADE$ gauge symmetry.   This is essentially clear
because the worldsheet instantons which wrap only around the fiber
wrap around individual $\IP^1$s and thus reproduce the sum of ${\rm log}$
terms one for each $\IP^1$ which are in one-to-one correspondence
with the positive roots of the gauge group.

Now we come to the more interesting corrections
corresponding to the gauge theory instantons, which
correspond to world sheet instantons wrapping around the base.
Let us now see what is expected in the
field theory limit.  In this case we are taking the
base to be large, and in particular $q_b\rightarrow \epsilon^4,q_f\rightarrow
1-\epsilon a$.
This in particular means that in this limit the multi-cover
contributions of worldsheet instantons are suppressed, because
they are in the denominator in the combination $(1-q_b^nq_f^m)\rightarrow 1$.
Ignoring the multicover contribution allows us to write the $\partial_{t_f}^2
\cF$ in a simple way
$$-i\tau =\partial_{t_f}^2 \cF\rightarrow \sum_{n,m} d_{n,m} m^2 q_b^nq_f^m$$
In the field theory limit we expect that the instanton number $n$
contribution to $\tau$ goes as $1/a^{4n}$.  Let us see
how such a behavior may emerge from the above sum.   In the limit
$q_f\rightarrow 1$ only the asymptotic growth of the
above sum is relevant.  Let us assume that $d_{n,m}\sim \gamma_n m^{\alpha
(n)}$
for large $m$.  Then the leading contribution to $\tau$ from
gauge theory instanton number $n$ gives a term proportional to
$$q_b^n\sum m^{2+\alpha(n)} q_f^m\propto {q_b^n\over (1-q_f)^{3+\alpha(n)}}
\propto {\epsilon^{4n}\over (\epsilon a)^{3+\alpha(n)}}$$
we thus expect, in order to obtain a non-trivial
field theory limit as $\epsilon \rightarrow 0$,
 that $\alpha(n)=4n-3$, i.e. we expect that
\eqn\gro{d_{n,m}\sim \gamma_n m^{4n-3}}
 for fixed $n$ in the limit of large $m$.  Moreover we expect
that $\gamma_n$ should not depend on how we realize the
gauge theory system, i.e. we should get the same result
for all ${\bf F}_n$ (up to some trivial overall rescaling
of $a$).

Let us inspect the number of instantons for the first few cases,
using the local mirror symmetry discussed in section 4.
For gauge theory instanton number one, which
corresponds to worldsheet instantons wrapping
once around the base and arbitrary number of times around
the fiber the numbers grows linearly.  For example we have
$$
d^{F_0}_{1,m}=-(2 m+2),\ \
d^{F_1}_{1,m}=2m+1,\ \
d^{F_2}_{1,m}=-2 m.$$

This is in accordance with the expected growth given in \gro .
Also notice that the coefficient of the growth $\gamma_1$ is the
same in all these cases (the fact that they differ by a sign
for odd instanton numbers for $F_1$ is a check on the relation
of $F_1$ with a discrete theta angle discussed in \ref\kdv{S. Katz,
M. Douglas and C. Vafa, hep-th/9609071}).
At $n=2$ one observes
$d^{F_1}_{2,0}=d^{F_2}_{2,0}=0$ and
$d^{F_0}_{2,i}=d^{F_1}_{2,i+1}=d^{F_2}_{2,i+1}$ where the number for
$d^{F_0}$ appear in the following table. The numbers for $F_0$
and $F_2$ for higher $m$ are just shifted
$d^{F_2}_{n,m}=d^{F_0}_{n,m-n-1}$, with $d_{n,m}^{F_2}=0$ for
$m-n-1<0$. The numbers for $F_1$ for higher $n$
are not reported. They differ from the $F_0$ and $F_2$ cases
but have a similar growth.

$$
\vbox{\offinterlineskip\tabskip=0pt
\halign{\strut
\vrule#&
{}~\hfil$#$~&
\vrule#&
{}~\hfil$#$~&
{}~\hfil$#$~&
{}~\hfil$#$~&
{}~\hfil$#$~&
{}~\hfil$#$~&
{}~\hfil$#$~&
\vrule$#$\cr
\noalign{\hrule}
& m &&d^{F_0}_{0,m} &  d^{F_0}_{1,m}&d^{F_0}_{2,m}&d^{F_0}_{3,m}
&d^{F_0}_{4,m}&\ldots & \cr
\noalign{\hrule}
&1 &&-2&-4&  -6     &       -8&         -10&&\cr
&2 && 0 &-6&  -32    &     -110&        -288&&\cr
&3 && 0 &-8&  -110   &     -756&       -3556&&\cr
&4 && 0 &-10& -288   &    -3556&      -27264&&\cr
&5 && 0 &-12& -644   &   -13072&     -153324&&\cr
&6 && 0 &-14& -1280  &   -40338&     -690400&&\cr
&7 && 0 &-16& -2340  &  -109120&    -2627482&&\cr
&8 && 0 &-18& -4000  &  -266266&    -8757888&&\cr
&9 && 0 &-20& -6490  &  -597888&   -26216372&&\cr
&10&& 0 &-22& -10080 & -1253538&   -71783040&&\cr
&11&& 0 &-24& -15106 & -2481024&  -182298480&&\cr
&12&& 0 &-26& -21952 & -4675050&  -434054144&&\cr
&13&& 0 &-28& -31080 & -8443424&  -977304976&&\cr
&14&& 0 &-30& -43008 &-14695208& -2095334784&&\cr
&15&& 0 &-32& -58344 &-24755858&     &&\cr
&16&& 0 &-34& -77760 &    &            &&\cr
&\vdots&&&   &        &         &            &&\cr
\noalign{\hrule}
&\gamma_n&& 0 &2  & {1\over 12}& {1\over 1890}&{113\over 119750400}&&\cr
\noalign{\hrule}
}
\hrule}$$
 The growth in the instanton number is in  accordance
with expectations based on field theory \gro .  In the limit we are discussing
the precise relation of
$\gamma_n$ with the corrected prepotential is obtained as follows.  In the
$\epsilon \rightarrow 0$ limit we have
$$\partial_{a}^2 \cF= 2 \log a - \sum_{n=1}^\infty \c_n
{(4 n-1)!\over a^{4n}}.$$

By comparison with the $N=2$ gauge coupling constant
$$\tau={4 i\over \pi}\left(\log\left(\tilde a\over \Lambda\right)+c-{1\over 8}
\sum_{n=1}^\infty (4n-2)(4n-1)
\left(\Lambda\over \tilde a\right)^{4 n}\cF_n \right).$$
we can fix the proportionality constant to be $a\sim
2\cdot 2^{1/4} \tilde a$ (note that
the overall rescaling of $\cF$ is not physically
relevant). The space-time instantons
and the asymptotic growth of the worldsheet instantons are related by
$$\c_n = {2^{3(3 n-1)}\over (4n-3)!} \cF_n.$$
The $\cF_n$ can be readily calculated since $\cF(a)$ is completely
determined by the periods $a(u)=\int_a \lambda\ $,
$\partial_a \cF(a)=a_D=\int_b \lambda \ $ of  the meromorphic form
$\lambda= {i \sqrt{2}\over 4 \pi} 2 x^2 {d x\over y}$ over the
Seiberg-Witten curve \swii
\eqn\swc{y^2=(x^2-u)^2-\Lambda^4.}
The first few are $\cF_n={1\over 2^5},{5\over 2^{14}},{3\over 2^{18}},
{1469\over 2^{31}}\ldots$ for $n=1,2,\ldots$ .
The prediction that the number of worldsheet instantons
grow asymptotically as $d_{n,m} \sim 2 m,
{1 \over 12} m^5,
{1\over 1890} m^9,
{113  \over 119750400}m^{13},\ldots$ for $n=1,2,\ldots$ is in good
agreement with
the data in table 1.
Below we will also establish this fact to all
orders both geometrically as well as
by showing that the differential equation which governs
$\cF$ goes in a particular limit \kklmv\ to the differential
equation for $a_D,a$ of the Yang-Mills system \swc .

\subsec{Embedding of the Seiberg-Witten curve.}
Before we consider the Picard-Fuchs
equations which govern $\cF (a)$ let us see the very simple way in which
the Seiberg-Witten curves arise from local mirror symmetry.
In the previous section we discussed the $F_2$ geometric construction
in detail.
As discussed in the previous section,
the mirror geometry  is given by the Riemann surface
\eqn\suII{P= a_1 z + a_2 {1\over z} + b_2 + b_1  t + b_0{ t}^2=0.}
The good algebraic coordinates, which are invariant under scaling
w.r.t the charge vectors are
$$z_b={a_1 a_2\over b_2^2}, \quad z_f={b_0 b_2\over b_1^2}$$
and we will set $a_1=a_2=b_0=1$ in the following.

Now the only thing we need to know is that the base becomes
large with $z_b\sim q_b\sim e^{-t_b}\sim \epsilon^4$, which
implies that $b_2\sim {1\over \epsilon^2}$.
What remains to complete
the moduli identification and taking the limit to the
$SU(2)$ elliptic curve is to bring  \suII\ in the
form used in \ref\gkmw{A. Gorskii, I. Krichever, A. Marshakov, A.
Mironov and A. Morozov, \plt355(1995) 466,
H. Itoyama and A. Morozov,
hep-th/9511126 and hep-th/9512161,
E. Martinec and N. Warner,\nup459(1996) 97}\klmvw\ by
getting rid of the next to leading
term in $t$ by
rescaling $ t=(\sqrt{2} x- {b_1\over 2})$. This leeads to
\eqn\swsuiinm{P=z+{1\over z}+2 (x^2-u)=0,}
with $u={1\over2}\left(b_2-{b_1^2\over 4}\right)$ required to be
finite. This means $2\epsilon^2 u:=\left(1-{1\over 4} {1\over z_f}\right)$
which gives the precise description of the  limit we are taking in
the good algebraic coordinates.
Identifying $z=y-(x^2-u)$ the equation
\swsuiinm\ can be brought to the form \swc .

Now for the Calabi-Yau threefold case with pure $SU(2)$
we have in our conventions the one-form \ttr\
$$\int \log (t) {d z\over z}.$$

Using the change of variable $ t=\sqrt{2} x-{1\over 2} b_1$ with
$b_1\sim \epsilon^{-1}$, in the limit
$\epsilon\rightarrow 0$ we get the period of
$$\int \log (x-\cO ({1\over \epsilon})) {dz\over z}=-\log(\cO (\epsilon))\int
{dz\over z}+\epsilon \int x{dz\over z}$$
Note that the residue of the first term around $z=0$ gives
the period associated with $S$ field and the next
term gives the periods of $a$ and $a_D$ which as expected
are proportional to $\epsilon$.

\subsec{Specialization of the Picard-Fuchs equations}
Even though what we showed above is sufficient
to prove that we obtain the expected $SU(2)$ field
theory result, it is helpful to recast the solutions
of the periods in terms of the Picard-Fuchs equations.
In fact
the prepotential of the theory is completely determined by the
system of Picard-Fuchs equations.

Due to the the large symmetries
in the toric representation of the local mirror geometry
the PFs are easy to derive. Together with the straightforward description
of the limit above this can be viewed as an effective way of deriving
the Picard-Fuchs system for the rigid $N=2$ Yang-Mills theories.

As an example of this consider the case
of $\IP^1\times \IP^1$ for which we expect to again obtain
in the field theory limit $SU(2)$ without matter.
Using the PF equations \fzero\  and substituting
$z_1\ra {1\over  4 x^2},
z_2\ra {y\over 4 }$
($
\t1\ra -{1\over 2} \tx,
\t2 \ra \ty$) one gets a system
with discriminant components given by the divisors $x=0,y=0$ and
$$\Delta = (1 -  x)^2 - x^2 y=0. $$

{}From the variable identification above it is clear that that the
$SU(2)$ point is at a tangential intersection of $\Delta=0$ with $y=0$
and the identification of variables in which we can take the
$\epsilon\rightarrow 0$ limit keeping a finite $u$ is forced on us.

Physically we identify $y=e^{-S}=(\epsilon)^4\Lambda^4 e^{-\hat S}$
and take the double scaling limit considered in \kklmv
$$\eqalign{
x_1& = \epsilon^2 u = (1-x)\cr
x_2& = {\Lambda^2 e^{-\hat S/2}\over u}={\sqrt{y}\over 1-x},}$$
to recover in the limit $\epsilon \rightarrow 0$ the pure $SU(2)$
$N=2$ YM at weak coupling. Mathematically $(x_1=0,x_2=0)$
describes the normal crossing of $y=0$ and an exceptional
divisor and provides variables in which the periods degenerates
at worst logarithmically at the boundary of the moduli space.

The indices of the system
$\cL_1(x_1,x_2),\cL_2(x_1,x_2)$ (i.e. the shift from an integer of
the power growth of the periods in algebraic coordinates)
at
$(x_1,x_2)=0$ are $(0,0)$ and $({1\over 2},0)$ and the
solutions are schematically of the form
\eqn\solPIxPI{\eqalign{
1\cr
\sqrt{x_1}(1+\cO(x_1))(1-{1\over 16} x_2-\ldots)&=
\epsilon a({\Lambda^2 e^{-\hat S/2}\over u})
(1+\cO(\epsilon^2 u))\cr
2\log(x_1 x_2)+2\log(1+x_2) &= -S+2 \log(1+\epsilon^2 u)\cr
\sqrt{x_1}(1+\cO(x_1))((1-{1\over 16} x_2-\ldots) log(x_2)+
\cO(x_2))=&\epsilon a_D
({\Lambda^2 e^{-\hat S/2}\over u})
(1+\cO(\epsilon^2 u)),}}
where $a(u)$ $a_D(u)$ are the periods of classical $Su(2)$
Yang-Mills theory. One can easily establish the occurrence of
$a,a_D$ to all orders in $x_1={\Lambda^2\over u}$,
by noticing that
$\cL_{sw}=\lim_{x_2\rightarrow 0}\sqrt{x_1\over x_2}
\cL_2(x_1,x_2) \sqrt{x_1 x_2}$ is the PF operator for
$a,a_D$ while $\lim_{x_2\rightarrow 0}
\sqrt{x_1\over x_2}\cL_2(x_1,x_2) \sqrt{x_1 x_2}$
vanishes on the second and third solution of
\solPIxPI.

As a check we wish to show that in the field theory
limit we obtain the same PF equations even if we started
with other systems which realize pure $SU(2)$.  In particular we
consider the
pure $SU(2)$ Yang-Mills limit of the Picard-Fuchs
systems for $F_1$ and $F_2$.
The fact that $F_2$ should work follows already
from the discussion in section 5.1.  However it would
also be useful to see how the discriminant
loci look even in this case.
 The story turns
out to be  very similar to the $F_0$ case. For $F_1$
one has the system \fone\ and by using the variables
 $z_b=y$, $z_f={1\over 4 x}$ we get the discriminant components
$x=0,y=0$ as well as
$$\Delta_1=(1-x)^2- x^2 y +9 x (y-{3\over 4} y^2)=0,\quad
\Delta_2=1-y=0.$$
 The tangency between
$y=0$ and $\Delta_1=0$  is resolved in the in the same way as before
$x_1 = \epsilon^2 u = (x-1)/2, \quad
x_2 = {\Lambda^2 e^{-\hat S/2}\over u}={\sqrt{-8 y}\over x-1}$
apart from a different rescaling which sets the $u$ parameter to
the same scale as before.
The indices of the solutions near $(x_1,x_2)=(0,0)$, the principal
structure of logarithm and the appearance of the Seiberg-Witten
periods are exactly the same as it is in \solPIxPI. The third period
can be summed up to give now
$$2 \log(x_1 x_2)+{1\over 2}\log(1+{x_2\over 2})=-S+{1\over 2}
\log(1+{\epsilon^2 u \over 2}).$$
Away from the $x_2 \ra 0$ limit the second and fourth periods
coincide for the $\IP^1\times \IP^1$- and $F_1$-case in many
coefficients but not completely.

The $F_2$ system \ftwo\
has discriminant components $z_b=0,z_f=0$ as well and
$$\Delta_1=(1-4 z_f)^2- 64 z_f^2 z_b=0,\quad
\Delta_2=1-4 z_f=0$$
After the variable change
$x_1 = \epsilon^2 u = (4 z_f-1)/2,\ \  x_2 =
{\Lambda^2 e^{-\hat S/2}/u}=2 \sqrt{z_b}/(4z_f-1)$
one recovers again the Seiberg-Witten theory for $x_1=0$.

\newsec{Generalizations}
So far we have mainly concentrated on one example, namely $SU(2)$ without
matter.  This however was just to illustrate the basic idea. Our
methods generalize to many other interesting cases.  In this section
we will first illustrate this in the context of some concrete examples,
in particular $SU(n+1)$ without matter and with one fundamental.  At
the end we will sketch the general idea and what is involved in
getting more general gauge groups with matter.

\subsec{$SU(n+1)$ without matter}

Geometrically we have an $A_n$ fiber over a base $\IP^1$.
The relevant K\"ahler parameters appearing in the
prepotential are therefore the volume of the base $t_b$ and
the individual fibers $t_{f_1},\ldots,t_{f_n}$ and we are
looking at the limit in which the size of the base becomes
large, while the one of the fibers are of the same order:
\eqn\scsun{
(1- q_{f_i})\sim \epsilon a_i, \qquad q_b\sim \epsilon^{2n+2}}
By the definition of the $A_n$ fiber we have $n$ K\"ahler
classes corresponding to the simple roots of $A_n$ and as many
$-2$ spheres as there are positive roots.
As discussed before, resuming up the instanton
contributions coming from these $-2$-spheres plus their multicovers
gives the perturbative contribution of the prepotential
\eqn\clsun{\cF_{1-loop}(a(u))={i\over 4 \pi}
\sum_{positive\atop roots\ \alpha} Z_\alpha^2
\log[Z_\alpha^2/\Lambda^2],}
where the $a_i$ parameterize vacuum expectation value of the
field in the Cartan subalgebra of the gauge group and
$Z_a=\vec q \cdot \vec a$.

The charge vectors defining the local toric variety $X$ describing
an $A_n$ fibration over $\IP^1$  are
\def\sp{\phantom -}
\eqn\suiiim{\eqalign{
v_b    &=
(1,1,   -2,\sp 0,\sp 0,\sp 0, \sp 0,\ldots,\sp 0,\sp 0,\sp 0,\sp 0)\cr
v_{f_1}&=
(0,0,\sp 1,  -2 ,\sp 1,\sp 0, \sp 0,\ldots,\sp 0,\sp 0,\sp 0,\sp 0)\cr
v_{f_2}&=
(0,0,\sp 0, \sp 1 , -2 ,\sp 1, \sp 0,\ldots,\sp 0,\sp 0,\sp 0,\sp 0)\cr
v_{f_3}&=
(0,0,\sp 0, \sp 0 ,\sp 1,  -2, \sp 1,\ldots,\sp 0,\sp 0,\sp 0,\sp 0)\cr
\vdots\ \  & \qquad \qquad \qquad \vdots  \cr
v_{f_{n-1}}&=
(0,0,\sp 0, \sp 0 ,\sp0,  \sp 0, \sp 0,\ldots,\sp 1,-2,\sp 1,\sp 0)\cr
v_{f_n}&=
(0,0,\sp 0, \sp 0 ,\sp 0,  \sp 0, \sp 0,\ldots,\sp 0,\sp 1,-2,\sp 1)}}
The $n+4$ columns of \suiiim\ were determined by the procedure described
earlier for $SU(2)$ using toric divisors.  In this case, the first two are the
$A_n$ fibers, the next is a divisor meeting the $A_n$ surface in a section
with self intersection $-2$ (very much like the divisor $x_3=0$ in the $F_2$
case).  The next $n$ divisors are the $n$ irreducible components of the
$A_n$ surface, and finally the last divisor meets the $A_n$ surface in a
disjoint section.

Local mirror symmetry associated to the columns of \suiiim\ the monomials
$z,{1\over z},1, t,\ldots t^{n+1}$ which satisfy the constraints required by
the charge vectors. The linear
constraint defining the family of curves is given by
\eqn\toricform{P=z+{1\over z}+ b_{n+1}+b_{n} t+\ldots b_1 t^{n}+b_0
t^{n+1}=0.}
The good coordinates are $z_b={1\over b_{n+1}^2}$,
$z_{f_{n-i}}={b_{i+1}b_{i-1}\over b_i^2}$,\ $i=1,\ldots,n$.
{}From \scsun\ we have $b_{n+1}\sim \epsilon^{-n-1}$. The scaling of the the
other
variables follows by shifting $t=(2^{1\over n+1} x -{b_1\over (n+1)})$
comparing with \gkmw
\eqn\standardform{P=z+{1\over z} + 2 W_{A_n}(x,\vec u)=0}
and requiring that all Weyl invariant parameters $u_i$ stay finite.
Especially that implies that $b_k\sim \epsilon^{-k}$ in leading order.
For the relation between the parameters $a_i$ used in \clsun\
and the Weyl invariant parameters $u_i$ in
$W_{A_n}(x,\vec u)$ see e.g. \ref\klt{A. Klemm, W. Lerche and S. Theisen,
Int. Journ. Mod. Phys. ${\underline {\rm A11}}$ (1996) 1929, hep-th/9505150}.

The  charge vectors $v_b,v_{f_1}\ldots,v_{f_n}$ are sufficient to write
down the relevant system of differential operators
$\cL_b,\cL_1$ and $\cL_k$ for $k=2,\ldots,n$
\eqn\genan{\eqalign{
\cL_b&=  \tb^2-z_b\prod_{i=0}^1(2\tb-\theta_{f_1}+i),\cr
\cL_1&=(\theta_{f_1}-2 \tb )(\theta_{f_1}-2\theta_{f_2})-z_{f_1}
\prod_{i=0}^1(2\theta_{f_1} -\theta_{f_2} +i),\cr
\cL_{k}&=
(\theta_{f_k}\! \! -\! 2\theta_{f_{k-1}}\!\!+\! \theta_{f_{k-2}})
(\theta_{f_k}\!\! - \! 2\theta_{f_{k+1}}\!\!+\! \theta_{f_{k+2}} )-
z_{f_k}
\prod_{i=0}^1
(2\theta_{f_k}\!\! -\! \theta_{f_{k-1}}\!\!-\! \theta_{f_{k+1}}+i),}}
where $\theta_{f_l}$ has to be omitted if $l$ out of the index range
$1<l<n$.
Further $n(n-1)/2$ second order differential operators follow by
considering combinations of the $v_{f_i}$,
which correspond to positive roots of $A_n$ and $n-1$ by
factorisation. At $z_b=z_{f_l}=0$ the system has a constant solution,
corresponding to the fact that the meromorphic form \oneform\
has a pole with non vanishing residue,
$n+1$ solutions starting with $\log(z_b)$, $\log(z_{f_i})$
and one solution which is quadratic in the logarithms.

Let us discuss the $SU(3)$ case in more detail. The system \genan\
has in this case besides the universal
discriminant $\Delta_b=1-4 z_b=0$ a principal
discriminant
\eqn\pdissuIII{\eqalign{
\Delta_p= &(1-4 z_{f_1}-4 z_{f_2}+18 z_{f_1}z_{f_2}-
27 z^2_{f_1} z^2_{f_2})^2-
\cr &8 z_{b} z_{f_1}^2(8-9 z_{f_2}(8-
3 z_{f_2}(7+4z_{f_1})+
6 z_{f_2}^2(2+9z_{f_1})+
81z_{f_2}^3z_{f_1}^2 (2 z_b-1)))=0.}}
The first line, which survives the $z_b\rightarrow 0$
limit, is recognized as the $SU(3)$ discriminant. It is
present if four points lie on an edge in a toric diagram, as it is
familiar e.g. from strong coupling gauge $SU(3)$
enhancements \kmp\km. Comparing \toricform \ with \standardform \ we have
\eqn\variableind{v= {2b_1^2\over 27}-{b_1 b_2\over 3}+b_3,\ \ \ \
   u=-{2^{1/3} b_1^2\over 3}+2^{1/3} b_2.}
In the limit $b_3\sim \epsilon^{-3}$ we get from this in leading orders
\eqn\limitsuIII{z_b={\Lambda^6 3^4\over 2^2} \epsilon^6,\ \ \
         z_{f_1}={1\over 3}+ {1\over 3^{2\over 3}} \epsilon^2 u,\ \ \
         z_{f_2}={1\over 3}+ {1\over 3^{2\over 3}} \epsilon^2 u+
                  3\epsilon^3 v,}
which when inserted in \pdissuIII\ reproduces of course in leading
order the rigid $SU(3)$ discriminant
$$(u^3-27(\Lambda^3+v)^2)(u^3-27(\Lambda^3-v)^2)\epsilon^{12} +
\cO(\epsilon^{13})$$
There are various choices of resolution variables to solve
the theory near $z_b=0$ $z_{f_1}= z_{f_2}={1\over 3}$  and to
derive from the $\cL_i$  the rigid $SU(N)$ operators, e.g.
$$x_1=(z_{f_2}-z_{f_1})\sim \epsilon^3 v,\ \ \
  x_2={\sqrt{z_{b}}\over (z_{f_2}-z_{f_1})}\sim {1\over v},\ \
  x_3={(z_{f_1}-{1\over 3})^{3/2}\over (z_{f_2}-z_{f_1})}
\sim\sqrt{u^3\over v^2}$$

\subsec{SU(n+1) with matter }

The generalization to the $A_n$ series with matter is straightforward.
The charge vectors for $A_n$ with one fundamental matter multiplet
($N_f=1$) are given by

$$\eqalign{
v_s    &=
(1, \sp 1, \sp 0,-2,\sp 0,\sp 0, \sp 0,\sp 0,\ldots,\sp 0,\sp 0,\sp 0,\sp 0)\cr
v_{f_1-E}&=
(0,-1,\sp 1, \sp 1  ,- 1,\sp 0, \sp 0,\sp 0, \ldots,\sp 0,\sp 0,\sp 0,\sp 0)\cr
v_{E}&=
(0,\sp 1, -1, \sp 0 , -1 ,\sp 1, \sp 0,\sp 0,\ldots,\sp 0,\sp 0,\sp 0,\sp 0)\cr
v_{f_2}&=
(0,\sp 0,\sp 0, \sp 0 ,\sp 1,  -2, \sp 1,\sp 0\ldots,\sp 0,\sp 0,\sp 0,\sp
0)
\cr
v_{f_3}&=
(0,\sp 0,\sp 0, \sp 0 ,\sp 1,  -2, \sp 1,\sp 0\ldots,\sp 0,\sp 0,\sp 0,\sp
0)\cr
\vdots\ \  & \qquad \qquad \qquad \vdots  \cr
v_{f_{n-1}}&=
(0,\sp 0,\sp 0, \sp 0 ,\sp0,  \sp 0, \sp 0,\sp 0,\ldots,\sp 1,-2,\sp 1,\sp
0)\cr
v_{f_n}&=
(0,\sp 0,\sp 0, \sp 0 ,\sp 0,  \sp 0, \sp 0, \sp 0 \ldots,\sp 0,\sp
1,-2,\sp 1)}$$

For $A_n=SU(n+1)$, this has $n+5$ columns and $n+2$ rows.
The $A_n$ surface has $n$ irreducible components, each of which is a ruled
surface over ${\bf P}^1$.  These components form a chain as dictated by the
$A_n$ Dynkin diagram; adjacent components meet along a section of each.
chain, the blowup being needed to obtain matter according to \kv.

The rows correspond to curves  $s,f_1-E,E,f_2,...,f_n$
where $s$ is the $-2$ curve in the $F_2$ at the end of the chain, $f_i$ are the
individual fibers, and $E$ is the exceptional divisor obtained by blowing up a
point of $F_2$ ($E$ lies in the blown up $F_2$ and meets the next surface in
the chain by choice).  The blowup splits a particular $f_1$
fiber into two pieces, one of which is $E$ and the other being $f_1-E$.

The columns correspond to divisors $f,f-E,E,s,E_1,...,E_n,H$,
where $f$ is a fiber of the entire $A_n$ surface (so that $f=\sum f_i$),
$E_i$ are the surfaces comprising the chain, and $H$ is a section of
the last surface in the chain which is disjoint from the other surfaces in
the chain.

{}From the charge vectors, we see that
we can therefore associate to the rows the monomials
$1/z,z,zt,1,t,t^2,...,t^{n+1}$,
so that we just have to add on a single term with a $zt$ to get the matter.

$N_f=2$ can be done similarly --- there is a symmetric choice of the blowup
(i.e.\ simply blow up a point on the last component of the chain which meets
the next to last component in the chain).
The chain has been chosen to start with a blown up $F_2$ at the end of the
chain, the blowup being needed to obtain matter according to \kv.

Let us discuss the $SU(2)$ case with one matter in more detail. In that
case we have the charge
vectors
\eqn\suiim{\eqalign{
           v_b&= (\sp 1,\sp 1,\sp 0,   -2,\sp 0, \sp 0),\cr
       v_{F-E}&= (\sp 0,   -1,\sp 1,\sp 1,   -1, \sp 0),\cr
           v_E&= (\sp 0,\sp 1,   -1,\sp 0,   -1, \sp 1).}}

The corresponding relations are fulfilled by the following
parameterization of the coordinates of the mirror
$z,1/z,t/z,1,t,t^2$ which is given again by the linear constraint
\eqn\lpsuIIm{P=a_1 z+ a_2 {1\over  z} + a_3 {t \over z}+b_2+b_1 t+b_0 t^2=0}
After the same change of variables as before
$t\rightarrow \sqrt{2}x -{1\over 2}b_1$ and setting $z=y-(x^2-u)$ we
arrive at the form \swii\ of the elliptic curve
\eqn\suIImsw{y^2=(x^2-u)^2-\Lambda^3(x+m)}
with the parameter identification
\eqn\parameter{
        u=-{1\over  2}(b_2-{1\over 4} b_1^2),\ \ \
\Lambda^3=\sqrt{2} a_3, \ \ \
\Lambda^3 m=(a_2-{1\over 2} a_3 b_1).}
This means that we have to scale $b_2\sim \epsilon^{-2}$,
$b_1\sim \epsilon^{-1}$ and $a_2\sim \epsilon^{-1}$
(we set $a_1=a_3=b_0=1$ in the following).
Expressed in the good algebraic coordinates $z_b={a_2/b_2^2}\ $,
$z_{F-E}=b_2/(a_2 b_1)$ and $z_E=a_2/b_1$ the limit is
$$ z_b\sim \epsilon^3,\ \ \
({1\over 4} {1\over z_E z_{F-E}} -1)\sim 2 \epsilon^2 u, \ \ \
(z_E-{1\over 2})\sim \Lambda^3 m \epsilon,$$
which is consistent with the expected limit of
 growing base and shrinking fiber.
It has also been checked that the PF system and the discriminant
reduce to the one of $N=2$ $SU(2)$ gauge theory with matter.

It is similarly straightforward to check that the $SU(n+1)$ theory with one
matter also reduces to the expected result from field theory.

\subsec{Main idea in generalizations}
It should be clear from the above examples that
whenever we can construct a quantum field theory
by `geometric engineering' of fibered $ADE$ singularities
with possibly some extra singularities and at the same
time find the PF equations for the dual mirror
we would have solved the Coulomb branch of the corresponding
quantum field theory.  This thus involves 2 basic steps:
1) Construction of the local model; 2) Finding the mirror periods.

To construct the local mirror $\hat{X}$ of a fibered $ADE$ singularity, we
first review what we have done in our earlier examples.  We
first found generators for the Mori cone of the $ADE$ surface.  Then we
identified certain divisors.  These divisors can be described in hindsight
without toric geometry.  There were two fibers (modified as appropriate if
there were blowups).  We also required a fiber for each of the irreducible
components of the $ADE$ surface.  We also required a divisor class for each
blowup.  Finally, we required two disjoint sections.

In this way, we can take any geometric $ADE$ singularity and any geometric
description of extra matter as explained in \kv\ and now calculate charge
vectors from the geometry without using toric geometry directly.  The
result will generally be a toric description of $X$.

In the best cases, the
field $\prod x_i$ will be neutral, so that $X$ is itself a toric variety.
Then our methods will go through unmodified.  We do know that this cannot
happen in all examples; but then we can typically
construct $X$ as a (noncompact)
hypersurface or complete intersection in a toric variety.
Then the usual geometric constructions
of mirror symmetry can be used to construct $\hat{X}$.

Once we have $\hat{X}$, we can deduce the Picard-Fuchs operators directly
from those of compact Calabi-Yau threefolds associated to reflexive
polyhedra which `enlarge' the theory.  The scaling behavior of the periods
which satisfy these equations is intermediate in complexity between the
cases considered above and those arising from reflexive polyhedra.  We are
optimistic that the scaling behavior will lead us to realize the periods
as periods of a meromorphic differential on a curve directly from these
techniques.

{\bf Acknowledgments: }
We would like to thank B. Greene, W. Lerche,
P. Mayr, D. Morrison, N. Warner and S.-T. Yau for valuable
discussions.

The research of SK was partially supported by
NSF grant DMS-9311386 and NSA grant MDA904-96-1-0021, that of
 AK was partially supported by the Clay Fund for
Mathematics, through the Department of Mathematics, Harvard University
and that of CV was supported in part by NSF grant
PHY-92-18167.

\listrefs

\bye